\pdfoutput=1

\documentclass[12pt,a4paper]{article}

\usepackage{ifthen} 
\newboolean{pdflatex}
\setboolean{pdflatex}{true} 

\newboolean{articletitles}
\setboolean{articletitles}{true} 

\newboolean{uprightparticles}
\setboolean{uprightparticles}{false} 

\newboolean{inbibliography}
\setboolean{inbibliography}{false} 

\usepackage[top=1in, bottom=1.25in, left=1in, right=1in]{geometry}

\usepackage{microtype}
\usepackage{lineno}  
\usepackage{xspace} 
\usepackage{caption} 

\usepackage{graphicx}  
\usepackage{color}
\usepackage{colortbl}
\graphicspath{{./figs/}} 

\usepackage{amsmath} 
\usepackage{amssymb}
\usepackage{amsfonts}
\usepackage{upgreek} 

\newcommand*\patchAmsMathEnvironmentForLineno[1]{%
\expandafter\let\csname old#1\expandafter\endcsname\csname #1\endcsname
\expandafter\let\csname oldend#1\expandafter\endcsname\csname
end#1\endcsname
 \renewenvironment{#1}%
   {\linenomath\csname old#1\endcsname}%
   {\csname oldend#1\endcsname\endlinenomath}%
}
\newcommand*\patchBothAmsMathEnvironmentsForLineno[1]{%
  \patchAmsMathEnvironmentForLineno{#1}%
  \patchAmsMathEnvironmentForLineno{#1*}%
}
\AtBeginDocument{%
\patchBothAmsMathEnvironmentsForLineno{equation}%
\patchBothAmsMathEnvironmentsForLineno{align}%
\patchBothAmsMathEnvironmentsForLineno{flalign}%
\patchBothAmsMathEnvironmentsForLineno{alignat}%
\patchBothAmsMathEnvironmentsForLineno{gather}%
\patchBothAmsMathEnvironmentsForLineno{multline}%
\patchBothAmsMathEnvironmentsForLineno{eqnarray}%
}

\usepackage{hyperref}    
\usepackage[all]{hypcap} 

\usepackage{xspace} 
\usepackage{upgreek}

\def\lhcb {\mbox{LHCb}\xspace}

\def\MagUp {\mbox{\em Mag\kern -0.05em Up}\xspace}

\ifthenelse{\boolean{uprightparticles}}%
{

 \def\Pmu         {\ensuremath{\upmu}\xspace}

 \def\Ppi         {\ensuremath{\uppi}\xspace}

 \def\Pphi        {\ensuremath{\upphi}\xspace}

 \def\Ppsi        {\ensuremath{\uppsi}\xspace}

 \def\PDelta      {\ensuremath{\Delta}\xspace}                 
 \def\PXi      {\ensuremath{\Xi}\xspace}                 
 \def\PLambda      {\ensuremath{\Lambda}\xspace}                 
 \def\PSigma      {\ensuremath{\Sigma}\xspace}                 
 \def\POmega      {\ensuremath{\Omega}\xspace}                 
 \def\PUpsilon      {\ensuremath{\Upsilon}\xspace}                 
 
 \def\PB      {\ensuremath{\mathrm{B}}\xspace}                 
                  
 \def\PD      {\ensuremath{\mathrm{D}}\xspace}

 \def\PJ      {\ensuremath{\mathrm{J}}\xspace}                 
 \def\PK      {\ensuremath{\mathrm{K}}\xspace}

 \def\Pb      {\ensuremath{\mathrm{b}}\xspace}                 
 \def\Pc      {\ensuremath{\mathrm{c}}\xspace}

 \def\Pi      {\ensuremath{\mathrm{i}}\xspace}

 \def\Pp      {\ensuremath{\mathrm{p}}\xspace}

 \def\Ps      {\ensuremath{\mathrm{s}}\xspace}

}
{

 \def\Pmu         {\ensuremath{\mu}\xspace}

 \def\Ppi         {\ensuremath{\pi}\xspace}

 \def\Pphi        {\ensuremath{\phi}\xspace}

 \def\Ppsi        {\ensuremath{\psi}\xspace}                 
                  
 \mathchardef\PDelta="7101
 \mathchardef\PXi="7104
 \mathchardef\PLambda="7103
 \mathchardef\PSigma="7106
 \mathchardef\POmega="710A
 \mathchardef\PUpsilon="7107
                  
 \def\PB      {\ensuremath{B}\xspace}                 
                  
 \def\PD      {\ensuremath{D}\xspace}

 \def\PJ      {\ensuremath{J}\xspace}                 
 \def\PK      {\ensuremath{K}\xspace}

 \def\Pb      {\ensuremath{b}\xspace}                 
 \def\Pc      {\ensuremath{c}\xspace}

 \def\Pi      {\ensuremath{i}\xspace}

 \def\Pp      {\ensuremath{p}\xspace}

 \def\Ps      {\ensuremath{s}\xspace}

}

\makeatletter
\ifcase \@ptsize \relax
  \newcommand{\miniscule}{\@setfontsize\miniscule{4}{5}}
\or
  \newcommand{\miniscule}{\@setfontsize\miniscule{5}{6}}
\or
  \newcommand{\miniscule}{\@setfontsize\miniscule{5}{6}}
\fi
\makeatother

\DeclareRobustCommand{\optbar}[1]{\shortstack{{\miniscule (\rule[.5ex]{1.25em}{.18mm})}
  \\ [-.7ex] $#1$}}

\def\mup        {{\ensuremath{\Pmu^+}}\xspace}
\def\mun        {{\ensuremath{\Pmu^-}}\xspace} 

\def\squark    {{\ensuremath{\Ps}}\xspace}

\def\cquark    {{\ensuremath{\Pc}}\xspace}

\def\bquark    {{\ensuremath{\Pb}}\xspace}

\def\pion   {{\ensuremath{\Ppi}}\xspace}

\def\pip    {{\ensuremath{\pion^+}}\xspace}
\def\pim    {{\ensuremath{\pion^-}}\xspace}

\def\kaon    {{\ensuremath{\PK}}\xspace}
  \def\Kbar    {{\kern 0.2em\overline{\kern -0.2em \PK}{}}\xspace}

\def\KorKbar    {\kern 0.18em\optbar{\kern -0.18em K}{}\xspace}

\def\Kp      {{\ensuremath{\kaon^+}}\xspace}
\def\Km      {{\ensuremath{\kaon^-}}\xspace}

\def\KS      {{\ensuremath{\kaon^0_{\mathrm{ \scriptscriptstyle S}}}}\xspace}

  \def\Dbar    {{\kern 0.2em\overline{\kern -0.2em \PD}{}}\xspace}

\def\DorDbar    {\kern 0.18em\optbar{\kern -0.18em D}{}\xspace}

\def\B       {{\ensuremath{\PB}}\xspace}
\def\Bbar    {{\ensuremath{\kern 0.18em\overline{\kern -0.18em \PB}{}}}\xspace}

\def\BorBbar    {\kern 0.18em\optbar{\kern -0.18em B}{}\xspace}
\def\Bz      {{\ensuremath{\B^0}}\xspace}
\def\Bzb     {{\ensuremath{\Bbar{}^0}}\xspace}
\def\Bu      {{\ensuremath{\B^+}}\xspace}

\def\Bp      {{\ensuremath{\Bu}}\xspace}

\def\Bs      {{\ensuremath{\B^0_\squark}}\xspace}

\def\jpsi     {{\ensuremath{{\PJ\mskip -3mu/\mskip -2mu\Ppsi\mskip 2mu}}}\xspace}
\def\psitwos  {{\ensuremath{\Ppsi{(2S)}}}\xspace}

  \def\Y#1S{\ensuremath{\PUpsilon{(#1S)}}\xspace}

\def\proton      {{\ensuremath{\Pp}}\xspace}
\def\antiproton  {{\ensuremath{\overline \proton}}\xspace}

\def\Lz          {{\ensuremath{\PLambda}}\xspace}
\def\Lbar        {{\ensuremath{\kern 0.1em\overline{\kern -0.1em\PLambda}}}\xspace}
\def\LorLbar    {\kern 0.18em\optbar{\kern -0.18em \PLambda}{}\xspace}

\def\Lb      {{\ensuremath{\Lz^0_\bquark}}\xspace}
\def\Lbbar   {{\ensuremath{\Lbar{}^0_\bquark}}\xspace}

\def\to                 {\ensuremath{\rightarrow}\xspace}

\def\CP                {{\ensuremath{C\!P}}\xspace}

\newcommand{\ACP}{{\ensuremath{{\mathcal{A}}^{\CP}}}\xspace}

\def\C#1      {\ensuremath{\mathcal{C}_{#1}}\xspace}                       
\def\Cp#1     {\ensuremath{\mathcal{C}_{#1}^{'}}\xspace}                    
\def\Ceff#1   {\ensuremath{\mathcal{C}_{#1}^{\mathrm{(eff)}}}\xspace}        
\def\Cpeff#1  {\ensuremath{\mathcal{C}_{#1}^{'\mathrm{(eff)}}}\xspace}       
\def\Ope#1    {\ensuremath{\mathcal{O}_{#1}}\xspace}                       
\def\Opep#1   {\ensuremath{\mathcal{O}_{#1}^{'}}\xspace}                    

\newcommand{\tev}{\ifthenelse{\boolean{inbibliography}}{\ensuremath{~T\kern -0.05em eV}\xspace}{\ensuremath{\mathrm{\,Te\kern -0.1em V}}}\xspace}
\newcommand{\gev}{\ensuremath{\mathrm{\,Ge\kern -0.1em V}}\xspace}
\newcommand{\mev}{\ensuremath{\mathrm{\,Me\kern -0.1em V}}\xspace}
\newcommand{\kev}{\ensuremath{\mathrm{\,ke\kern -0.1em V}}\xspace}
\newcommand{\ev}{\ensuremath{\mathrm{\,e\kern -0.1em V}}\xspace}
\newcommand{\gevc}{\ensuremath{{\mathrm{\,Ge\kern -0.1em V\!/}c}}\xspace}
\newcommand{\mevc}{\ensuremath{{\mathrm{\,Me\kern -0.1em V\!/}c}}\xspace}
\newcommand{\gevcc}{\ensuremath{{\mathrm{\,Ge\kern -0.1em V\!/}c^2}}\xspace}
\newcommand{\gevgevcccc}{\ensuremath{{\mathrm{\,Ge\kern -0.1em V^2\!/}c^4}}\xspace}
\newcommand{\mevcc}{\ensuremath{{\mathrm{\,Me\kern -0.1em V\!/}c^2}}\xspace}

\def\mum  {\ensuremath{{\,\upmu\mathrm{m}}}\xspace}

\def\invfb   {\ensuremath{\mbox{\,fb}^{-1}}\xspace}

\def\gsim{{~\raise.15em\hbox{$>$}\kern-.85em
          \lower.35em\hbox{$\sim$}~}\xspace}
\def\lsim{{~\raise.15em\hbox{$<$}\kern-.85em
          \lower.35em\hbox{$\sim$}~}\xspace}

\def\ptot       {\mbox{$p$}\xspace}
\def\pt         {\mbox{$p_{\mathrm{ T}}$}\xspace}

\def\evtgen     {\mbox{\textsc{EvtGen}}\xspace}

\def\geant      {\mbox{\textsc{Geant4}}\xspace}

\def\photos     {\mbox{\textsc{Photos}}\xspace}

\def\pythia     {\mbox{\textsc{Pythia}}\xspace}

\def\tell1  {TELL1\xspace}
\def\ukl1   {UKL1\xspace}

\newcommand{\eg}{\mbox{\itshape e.g.}\xspace}
\newcommand{\ie}{\mbox{\itshape i.e.}\xspace}

\usepackage{cite} 
\usepackage{mciteplus}

\usepackage{longtable} 
\usepackage{subfigure}

\def\LbTopKMuMu      {{\ensuremath{\Lb \to \proton\Km\mup\mun}}\xspace}
\def\LbTopKJPsi      {{\ensuremath{\Lb \to \proton\Km\jpsi}}\xspace}
\newcommand{\AT}{{\ensuremath{A_{\widehat{T}}}}\xspace}
\newcommand{\ATbar}{{\ensuremath{\kern 0.1em\overline{\kern -0.1em A}_{\widehat{T}}}}\xspace}
\newcommand{\aCPTodd}{{\ensuremath{a_\CP^{\widehat{T}\text{-odd}}}}\xspace}
\newcommand{\aPTodd}{{\ensuremath{a_P^{\widehat{T}\text{-odd}}}}\xspace}

\newcommand{\CT}{{\ensuremath{C_{\widehat{T}}}}\xspace}
\newcommand{\CTbar}{{\ensuremath{\kern 0.1em\overline{\kern -0.1em C}_{\widehat{T}}}}\xspace}

\def\LK        {\ensuremath{\Lb \to \proton\Km\mup\mun}\xspace}

\def\LbarK     {\ensuremath{\Lbbar \to  \antiproton\Kp\mun\mup  }\xspace}

\def\LKJ       {\ensuremath{\Lb \to \proton\Km \jpsi}\xspace}
\def\LpiJ       {\ensuremath{\Lb \to \proton \pim  \jpsi}\xspace}

\def\BsKKmm    {\ensuremath{\Bs\to \Kp\Km\mu^-\mu^+}\xspace}
\def\BdKpimm   {\ensuremath{\Bzb\to \Km \pip\mup\mun}\xspace}

\def\CPV       {\ensuremath{\rm CPV}\xspace}

\def\DACP      {{\ensuremath{\Delta\mathcal{A}_{\CP}}}\xspace}
\def\ACP      {\ensuremath{a_\CP^{\widehat{T}\text{-odd}}}\xspace}
\newcommand{\Araw}{\ensuremath{\mathcal{A}_{\text{raw}}}\xspace}

\usepackage{feynmp,scalerel,ifpdf}

\ifpdf
\fi

\makeatletter
\def\endfmffile{%
  \fmfcmd{\p@rcent\space the end.^^J%
          end.^^J%
          endinput;}%
  \if@fmfio
    \immediate\closeout\@outfmf
  \fi
  \IfFileExists{\thefmffile.mp}{\immediate\write18{mpost \thefmffile}}{}
  \let\thefmffile\relax
}
\makeatother

\newsavebox\feynLbpKccbar
\newsavebox\feynLbpKccbarpenguin
\newsavebox\feynLbpKmumupenguin
\newsavebox\feynLbpKmumubox
\newlength\tmplength

\begin{document}

\setlength{\unitlength}{1mm}


\savebox\feynLbpKmumupenguin{%
\begin{fmffile}{feynLbpKmumupenguin}
\begin{fmfgraph*}(36,36)
\fmfpen{thick}
\fmfstraight
\fmfleft{ld,lu,lb,ln1,ln2,ln3,ln4}
\fmfright{rd,ru1,ru2,rubar,rs,rmup,rmum}
\fmfv{label=$V^*_{bq}$,label.angle=150}{vW1}
\fmfv{label=$V_{qs}$,label.angle=-60}{vW2}
\fmf{fermion}{ld,rd}
\fmf{fermion}{lu,ru1}
\fmf{plain,right}{rubar,ru2}
\fmf{fermion}{lb,vW1}
\fmf{fermion,label=$q$,label.side=right}{vW1,vW2}
\fmf{fermion}{vW2,rs}
\fmffreeze
\fmf{photon,label=$W^-$,left}{vW1,vW2}
\fmfforce{0.5w,0.682h}{vgamma1}
\fmf{fermion}{rmup,vgamma2,rmum}
\fmf{photon,label=$\gamma$,label.side=left}{vgamma1,vgamma2}
\fmfv{label=$d$,label.angle=180}{ld}
\fmfv{label=$u$,label.angle=180}{lu}
\fmfv{label=$b$,label.angle=180}{lb}
\fmfv{label=$d$,label.angle=0}{rd}
\fmfv{label=$u$,label.angle=0}{ru1}
\fmfv{label=$u$,label.angle=0}{ru2}
\fmfv{label=$\overline{u}$,label.angle=0}{rubar}
\fmfv{label=$s$,label.angle=0}{rs}
\fmfv{label=$\mu^+$,label.angle=0}{rmup}
\fmfv{label=$\mu^-$,label.angle=0}{rmum}
\end{fmfgraph*}
\end{fmffile}
}

\savebox\feynLbpKmumubox{%
\begin{fmffile}{feynLbpKmumubox}
\begin{fmfgraph*}(36,36)
\fmfpen{thick}
\fmfstraight
\fmfleft{ld,lu,lb,ln1,ln2,ln3,ln4}
\fmfright{rd,ru1,ru2,rubar,rs,rmup,rmum}
\fmfv{label=$V^*_{bq}$,label.angle=150}{vW1}
\fmfv{label=$V_{qs}$,label.angle=-60}{vW2}
\fmf{fermion}{ld,rd}
\fmf{fermion}{lu,ru1}
\fmf{plain,right}{rubar,ru2}
\fmf{fermion}{lb,vW1}
\fmf{fermion,label=$q$,label.side=right}{vW1,vW2}
\fmf{fermion}{vW2,rs}
\fmffreeze
\fmfforce{0.43w,1h}{vW3}
\fmfforce{0.74w,0.833h}{vW4}
\fmf{photon,label=$W^-$,label.side=left}{vW1,vW3}
\fmf{photon,label=$W^+$,label.side=left,label.dist=2}{vW2,vW4}
\fmf{fermion}{vW3,rmum}
\fmf{fermion}{rmup,vW4}
\fmf{fermion,label=$\nu_{\mu}$,label.side=left,label.dist=2}{vW4,vW3}
\fmfv{label=$d$,label.angle=180}{ld}
\fmfv{label=$u$,label.angle=180}{lu}
\fmfv{label=$b$,label.angle=180}{lb}
\fmfv{label=$d$,label.angle=0}{rd}
\fmfv{label=$u$,label.angle=0}{ru1}
\fmfv{label=$u$,label.angle=0}{ru2}
\fmfv{label=$\overline{u}$,label.angle=0}{rubar}
\fmfv{label=$s$,label.angle=0}{rs}
\fmfv{label=$\mu^+$,label.angle=0}{rmup}
\fmfv{label=$\mu^-$,label.angle=0}{rmum}
\end{fmfgraph*}
\end{fmffile}
}

\renewcommand{\thefootnote}{\fnsymbol{footnote}}
\setcounter{footnote}{1}


\begin{titlepage}
\pagenumbering{roman}

\vspace*{-1.5cm}
\centerline{\large EUROPEAN ORGANIZATION FOR NUCLEAR RESEARCH (CERN)}
\vspace*{1.5cm}
\noindent
\begin{tabular*}{\linewidth}{lc@{\extracolsep{\fill}}r@{\extracolsep{0pt}}}
\ifthenelse{\boolean{pdflatex}}
{\vspace*{-2.7cm}\mbox{\!\!\!\includegraphics[width=.14\textwidth]{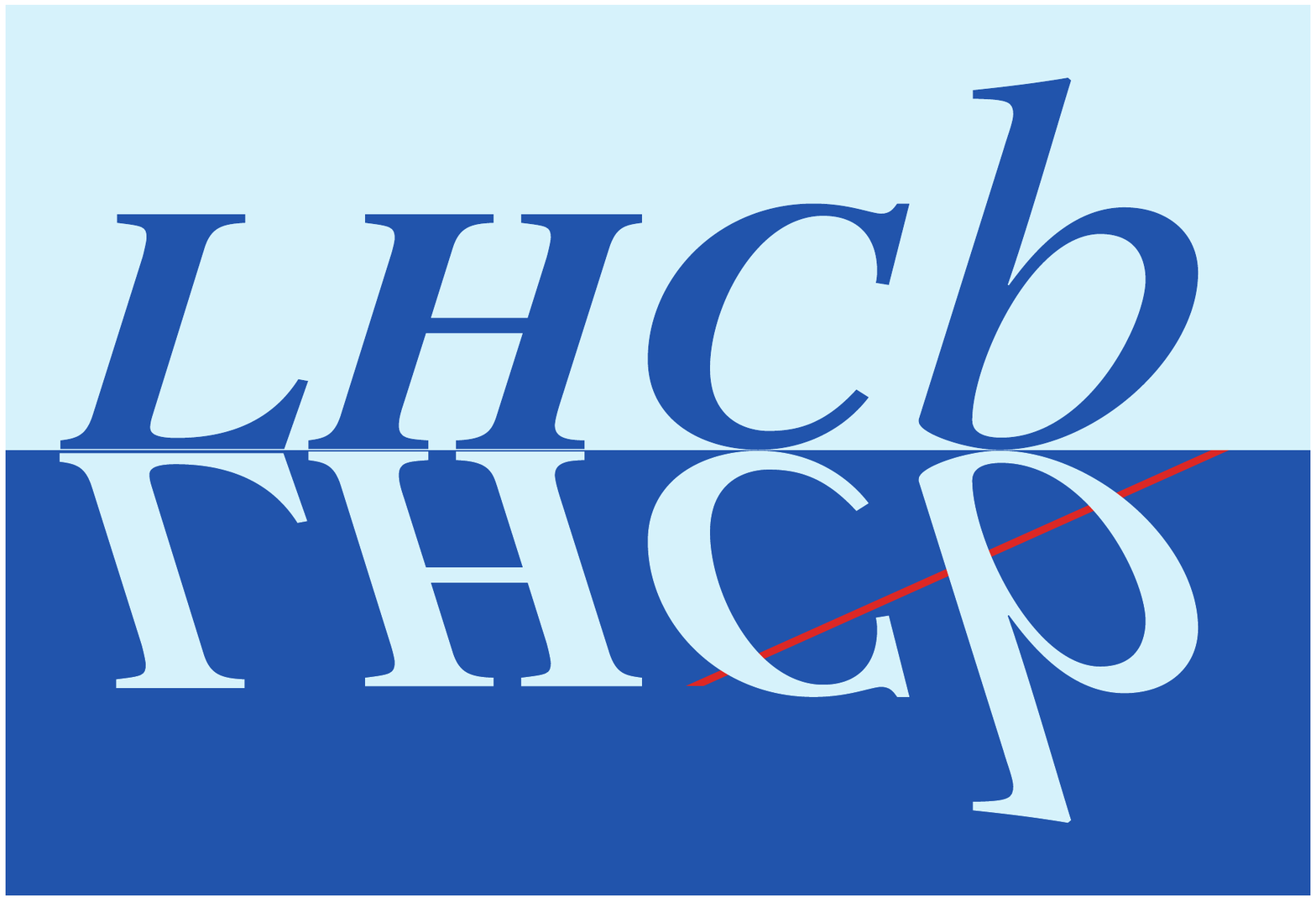}} & &}%
{\vspace*{-1.2cm}\mbox{\!\!\!\includegraphics[width=.12\textwidth]{lhcb-logo.eps}} & &}%
\\
 & & CERN-EP-2017-032 \\  
 & & LHCb-PAPER-2016-059 \\  
 & & \today \\ 
 & & \\
\end{tabular*}

\vspace*{2.8cm}

{\normalfont\bfseries\boldmath\huge
\begin{center}
Observation of the decay \LbTopKMuMu and a search for \CP violation
\end{center}
}

\vspace*{1.2cm}

\begin{center}
The LHCb collaboration\footnote{Authors are listed at the end of this paper.}
\end{center}

\vspace{\fill}

\begin{abstract}
  \noindent
A search for \CP violation in the decay \LbTopKMuMu is presented.
This decay is mediated by flavour-changing neutral-current transitions in the Standard Model and is potentially sensitive to new sources of \CP violation.
The study is based on a data sample of proton-proton collisions recorded with the \lhcb experiment, corresponding to an integrated luminosity of $3\invfb$.
The \LK decay is observed for the first time, and
two observables that are sensitive to different manifestations of \CP violation are measured, \mbox{$\DACP \equiv \mathcal{A}_{\CP}(\LbTopKMuMu)-\mathcal{A}_{\CP}(\LbTopKJPsi)$} and \aCPTodd, where the latter is based on asymmetries in the angle between the $\mup \mun$ and $p\Km$ decay planes. These are measured to be
\vspace{-0.1cm}
\begin{align*}
\DACP &= (-3.5\; \pm 5.0\;(\mathrm{stat})\; \pm 0.2\;(\mathrm{syst}))\times 10^{-2},\\
\aCPTodd &= (\phantom{+}1.2 \;\pm 5.0\;(\mathrm{stat}) \;\pm 0.7\;(\mathrm{syst}))\times 10^{-2},
\end{align*}
and no evidence for \CP violation is found.
  
\end{abstract}

\vspace*{2.0cm}

\begin{center}
  Published in JHEP 06 (2017) 108 
\end{center}

\vspace{\fill}

{\footnotesize 
\centerline{\copyright~CERN on behalf of the \lhcb collaboration, licence \href{http://creativecommons.org/licenses/by/4.0/}{CC-BY-4.0}.}}
\vspace*{2mm}

\end{titlepage}


\newpage
\setcounter{page}{2}
\mbox{~}
%
%
%
%

\cleardoublepage


\renewcommand{\thefootnote}{\arabic{footnote}}
\setcounter{footnote}{0}



\pagestyle{plain} 
\setcounter{page}{1}
\pagenumbering{arabic}


%

\section{Introduction}
\label{sec:introduction}

The phenomenon of \CP violation (\CPV), related to the difference in behaviour between matter and antimatter, remains an intriguing topic more than fifty years after its discovery in the neutral kaon system~\cite{PhysRevLett.13.138}. Within the Standard Model of particle physics (SM), \CPV is incorporated by a single, irreducible weak phase in the \mbox{Cabibbo-Kobayashi-Maskawa (CKM)} quark mixing matrix~\cite{1963PhRvL..10..531C,1973PThPh..49..652K}. However, the amount of \CPV in the SM is insufficient to explain the observed level of matter-antimatter asymmetry in the Universe~\cite{1991SvPhU..34..392S,Gavela:1993ts,Gavela:1994dt}. Therefore, new sources of \CPV beyond the SM are expected to exist. Experimental observations of \CPV remain confined to the $B$- and $K$-meson systems. Recently, the first evidence for \CPV in $\Lb\to p\pi^-\pi^+\pi^-$ was found at the level of $3.3$ standard deviations~\cite{LHCb-PAPER-2016-030} and a systematic study of \CPV in beauty baryon decays has now begun.

Among dedicated heavy-flavour physics experiments, the LHCb detector~\cite{Alves:2008zz} is unique in having access to a wide range of decay modes of numerous $b$-hadron species. Beauty baryons are produced copiously at the LHC, and within the LHCb detector acceptance the production ratio of $\Bz:\Lb:\Bs$ particles is approximately $4:2:1$~\cite{Aaij:2011jp}.
The \lhcb collaboration has previously searched for \CPV in \LpiJ and \LKJ decays~\cite{LHCb-PAPER-2014-020}, as well as in charmless $\Lb\to p\KS\pi^-$, $\Lb\to\Lz\phi$ and $\Lb\to\Lz h^ +h^-$ transitions~\cite{LHCb-PAPER-2013-061,LHCb-PAPER-2016-002,LHCb-PAPER-2016-004}.
\begin{figure}[h]
\centering
\includegraphics[scale=1.]{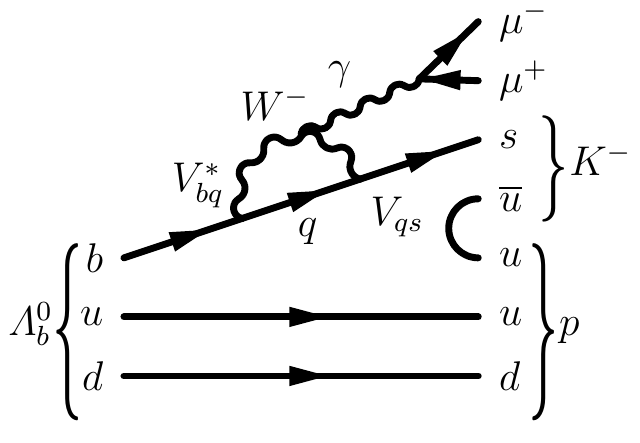}
\includegraphics[scale=1.]{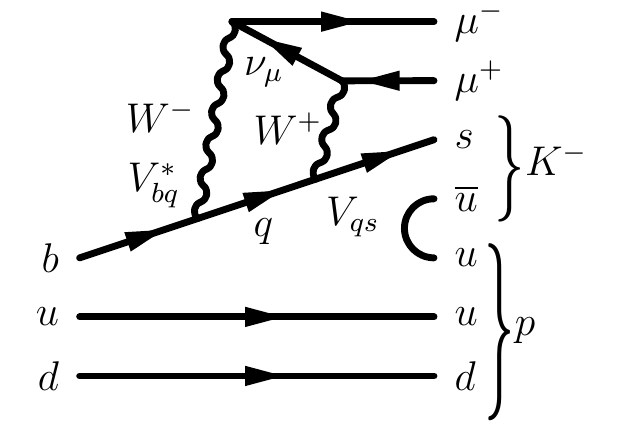}
\caption{Diagrams for the decay \LK, in which $V_{bq}$ and $V_{qs}$ are CKM matrix elements and $q$ represents one of the three up-type quarks $u$, $c$ or $t$, the $t$-quark contribution being dominant. The $u\overline{u}$ pairs originate from the hadronization process.\label{fey::LbpKmumu}}
\end{figure}

In this paper, a search for \CPV in the hitherto unobserved decay \LK is reported.\footnote{The inclusion of charge-conjugate processes is implied throughout this paper, unless stated otherwise.}
It is a flavour-changing neutral-current process with the underlying quark-level transition $b \to s \mup \mun$. The leading-order transition amplitudes in the SM are described by the loop diagrams shown in Fig.~\ref{fey::LbpKmumu}. In extensions to the SM, new heavy particles could contribute to the amplitudes with additional weak phases,
providing new sources of \CPV~\cite{Gauld:2013qja,Paracha:2014hca}. The limited amount of \CPV predicted for the decay \LK in the SM~\cite{Alok:2011gv,Paracha:2014hca}, following from the CKM matrix elements shown in Fig.~\ref{fey::LbpKmumu}, makes this decay particularly sensitive to \CPV effects from physics beyond the SM. 

\section{\boldmath\CP-odd observables}
\label{sec:cpv}
Two types of \CP-odd observables are studied in this paper. Following Refs.~\cite{Durieux:2015zwa,LHCb-PAPER-2016-030}, the differential rate of any pair of \CP-conjugate processes can be decomposed into four parts with definite even and odd transformation properties under the \CP and motion-reversal $\widehat{T}$ operators. Here, $\widehat{T}$ is the unitary operator that reverses both momentum and spin
three-vectors, to be distinguished from the antiunitary time-reversal operator $T$ which reverses initial and final states.

A $\widehat{T}$-even and \CP-odd asymmetry, $\mathcal{A}_{\CP}$, is related to the raw asymmetry $\mathcal{A}_{\text{raw}}$ of the observed decay candidates
\begin{equation}
\mathcal{A}_{\text{raw}} \equiv \frac{N(\LK)-N(\LbarK)}{N(\LK)+N(\LbarK)},
\end{equation}
via
\begin{equation}
\mathcal{A}_{\text{raw}} \approx \mathcal{A}_{\CP}(\LK) + \mathcal{A}_{\text{prod}}(\Lb) - \mathcal{A}_{\text{reco}}(K^+) + \mathcal{A}_{\text{reco}}(p)\label{eq::rawCP},
\end{equation}
where $\mathcal{A}_{\text{prod}}(\Lb)$ is the \Lb production asymmetry, due to the $pp$ initial state, and $\mathcal{A}_{\text{reco}}(K^+)$ and $\mathcal{A}_{\text{reco}}(p)$ are the reconstruction asymmetries for kaons and protons, mainly due to the different interaction cross-sections of particles and antiparticles with the detector material. By measuring the difference of raw asymmetries between the signal and the Cabibbo-favoured control mode $\LKJ(\to\mu^+\mu^-)$, the production and reconstruction asymmetries cancel to a good approximation. No significant \CPV is expected in the latter decay, since its amplitude is dominated by tree-level \CP-conserving diagrams, which leads to
\begin{equation}
\begin{split}
\DACP &\equiv \mathcal{A}_{\CP}(\LK) - \mathcal{A}_{\CP}(\LKJ)\\
&\approx\mathcal{A}_{\text{raw}}(\LK) -  \mathcal{A}_{\text{raw}}(\LKJ).
\end{split}
\label{eq::DACP}
\end{equation}
Imperfect cancellation in the production and reconstruction asymmetries can arise from differences in the kinematic distributions of the signal and control modes. A weighting procedure, discussed in Sec.~\ref{sec:asymmetries}, is applied to correct for this, with residual effects considered as a source of systematic uncertainty in Sec.~\ref{sec:systematics}.

\begin{figure}
\centering
\includegraphics[scale=0.35]{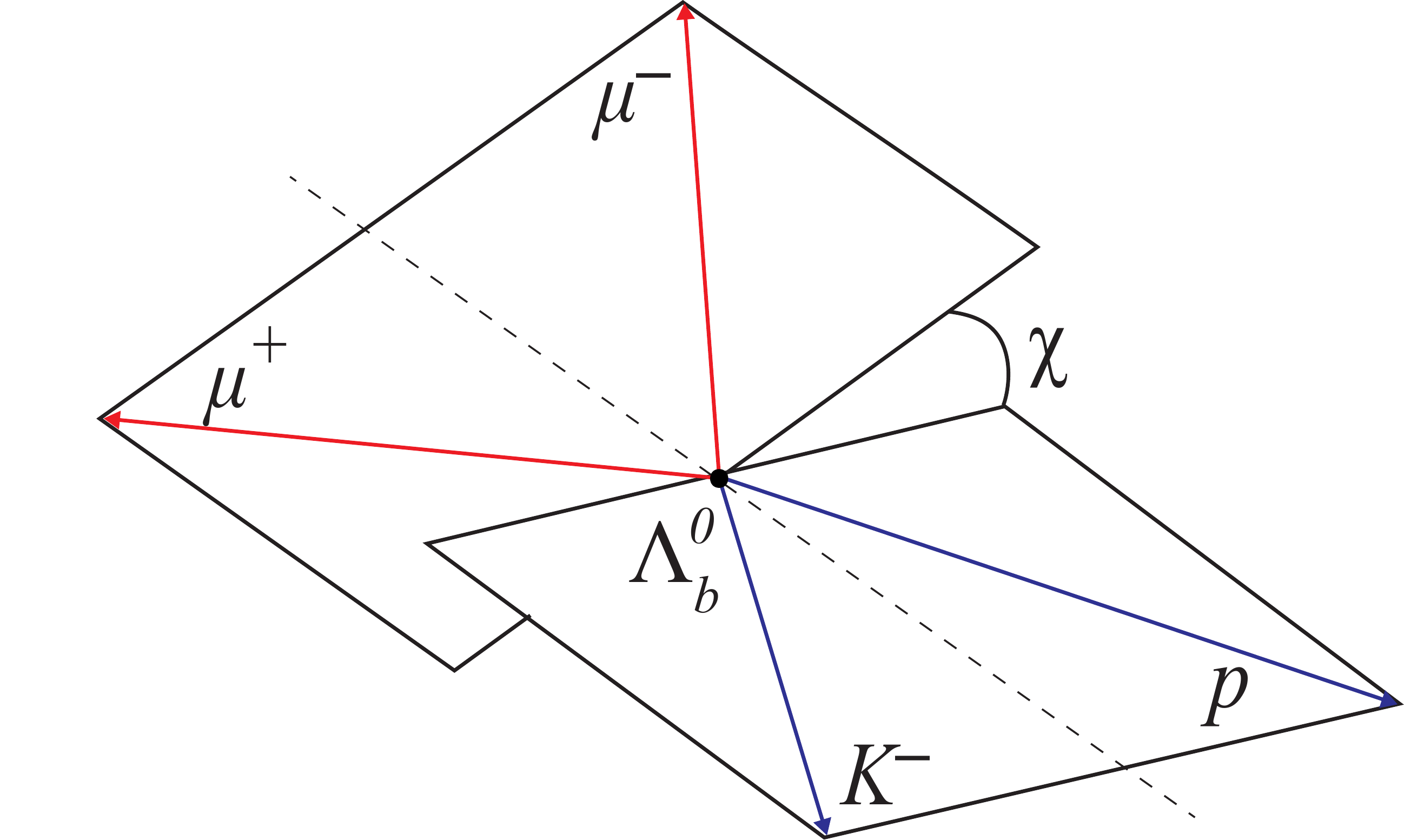}
\caption{Definition of the angle $\chi$ for \LK decays, in the \Lb rest frame.\label{fig:phiangle}}
\end{figure}

A pair of $\widehat{T}$-odd and $P$-odd observables, \AT and \ATbar, is obtained by defining the $\widehat{T}$-odd triple products of the final-state particle momenta in the \Lb rest frame
\begin{align}
\CT &\equiv \vec{p}_{\mu^+}\cdot (\vec{p}_{p}\times\vec{p}_{K^-}), \\
\CTbar &\equiv \vec{p}_{\mu^-}\cdot (\vec{p}_{\bar{p}}\times\vec{p}_{K^+}),\label{eq::TP}
\end{align}
and taking the asymmetries
\begin{equation}
\AT \equiv \frac{N(\CT > 0) - N(\CT < 0)}{N(\CT > 0) + N(\CT < 0)},
\hspace{0.5cm}
\ATbar \equiv \frac{\overline{N}(-\CTbar > 0) - \overline{N}(-\CTbar < 0)}{\overline{N}(-\CTbar > 0) + \overline{N}(-\CTbar < 0)},\label{eq::ATCP}
\end{equation}
where $N$($\overline{N}$) is the number of \Lb(\Lbbar) signal candidates. These asymmetries are measured from the angular distributions of the decay products, with $\CT$ being proportional to $\sin\chi$~\cite{Gronau:2011cf}, where $\chi$ is the angle between the decay planes of the $\mup \mun$ and $p\Km$ systems in the \mbox{\Lb rest frame}, as shown in Fig.~\ref{fig:phiangle}.

The observables \AT and \ATbar are $P$- and $\widehat{T}$-odd but are not sensitive to \CPV effects~\cite{Durieux:2015zwa}. Following Ref.~\cite{Gronau:2011cf}, \CP-odd and $P$-odd observables are defined as
\begin{equation}
\ACP \equiv \frac{1}{2} \left( \AT - \ATbar\right),\hspace{2cm} \aPTodd \equiv \frac{1}{2} \left( \AT + \ATbar\right),
\label{eq::ACPTodd}
\end{equation}
where a non-zero value of \ACP or \aPTodd would signal \CP or parity violation, respectively.
These observables are by construction largely insensitive to the \Lb production asymmetry and detector-induced charge asymmetries.

The observables \DACP and \ACP are sensitive to different manifestations of \CPV~\cite{Durieux:2015zwa}.
The \CP asymmetry $\mathcal{A}_{\CP}$ depends on the interference of $\widehat{T}$-even amplitudes, which can be written as $\mathcal{M}^e_i = a^e_i \exp\left[i(\delta^e_i+\phi_i^e)\right]$, where $\delta^e_i$ are \CP-even strong phases, $\CP(\delta^e_i) = \delta^e_i$, and $\phi_i^e$ are \CP-odd weak phases, $\CP(\phi_i^e)=-\phi_i^e$. This convention is such that all \CPV effects are encoded in the \CP-odd weak phases. The $\widehat{T}$-even and \CP-odd part of the differential rate turns out to be
\begin{equation}
\left.\frac{\rm d\Gamma}{\rm d\Phi} \right|^{\widehat{T}-{\rm even}}_{\CP-{\rm odd}} \propto a^e_1 a^e_2 \sin(\delta_1^e-\delta_2^e)\sin(\phi_1^e-\phi_2^e),
\end{equation}
where only two $\widehat{T}$-even amplitudes are considered for simplicity.
Therefore, $\mathcal{A}_{\CP}$ is enhanced when the strong phase difference between the two amplitudes is large.\\
On the other hand, \ACP depends on the interference between $\widehat{T}$-even and $\widehat{T}$-odd amplitudes, the latter written as $\mathcal{M}^o_i = a^o_i \exp\left[i(\delta^o_i+\phi_i^o+\pi/2)\right]$, following the same convention used for $\widehat{T}$-even amplitudes. The $\widehat{T}$-odd and \CP-odd part of the differential rate is therefore
\begin{equation}
\left.\frac{\rm d\Gamma}{\rm d\Phi} \right|^{\widehat{T}-{\rm odd}}_{\CP-{\rm odd}} \propto a^e_1 a^o_1 \cos(\delta_1^e-\delta_1^o)\sin(\phi_1^e-\phi_1^o),
\end{equation}
where one $\widehat{T}$-even and one $\widehat{T}$-odd amplitudes are considered for simplicity.
As a consequence, \ACP is enhanced when the strong phase difference vanishes. Furthermore, the observables \DACP and \ACP are sensitive to different types of \CPV effects from physics beyond the SM~\cite{Alok:2011gv}.

\section{Detector and simulation}
\label{sec:Detector}

The \lhcb detector~\cite{Alves:2008zz,LHCb-DP-2014-002} is a single-arm forward
spectrometer covering the \mbox{pseudorapidity} range $2<\eta <5$,
designed for the study of particles containing \bquark or \cquark
quarks. The detector includes a high-precision tracking system
consisting of a silicon-strip vertex detector surrounding the $pp$
interaction region, a large-area silicon-strip detector located
upstream of a dipole magnet with a bending power of about
$4{\mathrm{\,Tm}}$, and three stations of silicon-strip detectors and straw
drift tubes placed downstream of the magnet.
The tracking system provides a measurement of momentum, \ptot, of charged particles with
a relative uncertainty that varies from 0.5\% at low momentum to 1.0\% at 200\gevc.
The minimum distance of a track to a primary vertex (PV), the impact parameter (IP), 
is measured with a resolution of $(15+29/\pt)\mum$,
where \pt is the component of the momentum transverse to the beam, in\,\gevc.
Different types of charged hadrons are distinguished using information
from two ring-imaging Cherenkov detectors. 
Photons, electrons and hadrons are identified by a calorimeter system consisting of
scintillating-pad and preshower detectors, an electromagnetic
calorimeter and a hadronic calorimeter. Muons are identified by a
system composed of alternating layers of iron and multiwire
proportional chambers.
The online event selection is performed by a trigger~\cite{LHCb-DP-2012-004}, 
which consists of a hardware stage, based on information from the calorimeter and muon
systems, followed by a software stage, which applies a full event
reconstruction.

Simulated signal events are used to determine the effect of the detector geometry, trigger, reconstruction and selection on the angular distributions of the signal and \LKJ control sample. Additional simulated samples are used to estimate the contribution from specific background processes.
In the simulation, $pp$ collisions are generated using
\pythia~\cite{Sjostrand:2006za,Sjostrand:2007gs} 
 with a specific \lhcb
configuration~\cite{LHCb-PROC-2010-056}.  Decays of hadronic particles
are described by \evtgen~\cite{Lange:2001uf}, in which final-state
radiation is generated using \photos~\cite{Golonka:2005pn}. The
interaction of the generated particles with the detector, and its response,
are implemented using the \geant
toolkit~\cite{Allison:2006ve}, as described in
Ref.~\cite{LHCb-PROC-2011-006}.

\section{Selection of signal candidates}
\label{sec:sel}

The present analysis is performed using proton-proton collision data corresponding to \mbox{$1$ and} $2\invfb$ of integrated luminosity, collected with the LHCb detector in 2011 and 2012, at centre-of-mass energies of 7 and 8\tev, respectively. The \LK candidates are reconstructed from a proton, a kaon and two muon candidates originating from a common vertex, and are selected using information from the particle identification system. The \Lb flavour is determined from the charge of the kaon candidate, \ie \hspace{-4pt} \Lb for negative and \Lbbar for positive kaons.
%
Only candidates with reconstructed invariant mass, $m(pK^-\mu^+\mu^-)$, in the range $[5350,6000]\mevcc$ and a $pK^-$ invariant mass, $m(pK^-)$, below 2350\mevcc are retained, with the latter requirement being applied to reduce the combinatorial background contribution. The spectrum in the dimuon mass squared, $q^2$, is considered, excluding the resonance regions
$q^2 \in [0.98,1.10]$, $[8.0,11.0]$ and $[12.5,15.0]{\mathrm{\,Ge\kern -0.1em V^2\!/}c^4}$ that correspond to the masses of the $\Pphi(1020)$, \jpsi, and \psitwos mesons, respectively.

Several background contributions from exclusive decays are identified and rejected.
These are $\BsKKmm$ and $\BdKpimm$ decays, in which a kaon or a pion is
misidentified as a proton, and \LK decays, in which proton and kaon assignments are interchanged.
Background also arises from $\Lb\to p\Km \jpsi$ and $\Lb\to p \Km \psitwos$ decays in which a muon is misidentified as a kaon and the kaon as a muon.
These components are effectively eliminated by tightened particle identification
requirements combined with selection criteria on invariant masses
calculated under the appropriate mass hypothesis (\eg assigning the kaon mass to
the candidate proton to identify possible \BsKKmm background decays).
After these requirements the background contribution from the above decays is negligible. No indication of other specific background decays is observed.
The remaining combinatorial background is suppressed by means of a boosted decision tree (BDT)
classifier~\cite{Breiman,Roe} with an adaptive boosting algorithm \cite{AdaBoost}. The BDT is constructed from variables that discriminate between signal and background, based on their kinematic, topological and particle
identification properties, as well as the isolation of the final-state tracks~\cite{Aaij:2011rja,Aaij:2015bfa}. 
Simulated \LK events in which the decay products are uniformly distributed in phase space are used as the signal training sample and a correction for known differences between data and simulation is applied.
Candidates from data in the high mass region, $m(p\Km\mup\mun)>5800\mevcc$, are used as the background training sample and then removed from the window of the mass fit described below. After optimisation of the significance, $S/\sqrt{S+B}$, where $S$ and $B$ are the number of signal and background candidates in the region $m(pK^-\mu^+\mu^-) \in [5400,5800]\mevcc$, the BDT classifier retains only $0.14\%$ of the combinatorial background candidates, with a signal efficiency of $51\%$. Events in which more than one \Lb candidate survives the selection constitute less than $1\%$ of the sample and all candidates are retained; the systematic uncertainty associated with this is negligible. The identical selection is applied to the control-mode \LKJ, except that the dimuon squared mass is required to be in the range $[9.0,10.5]{\mathrm{\,Ge\kern -0.1em V^2\!/}c^4}$.

\section{Asymmetry measurements}
\label{sec:asymmetries}

For the \DACP measurement, the data are divided into two subsamples according to the \Lb flavour. For the measurements of the triple-product asymmetries, four subsamples are defined by the combination of the \Lb flavour and the sign of \CT(or \CTbar for \Lbbar). The reconstruction efficiencies are studied with simulated events and are found to be equal for all subsamples.

The observable \DACP can be sensitive to kinematic differences between the signal and control-mode decays that affect the cancellation of the detection asymmetries in Eq.~\ref{eq::DACP}. This is taken into account by assigning a weight to each \LKJ candidate such that the resulting proton and kaon momentum distributions match those of the signal \LK decays. These weights are determined from simulation samples for the signal and control modes. No such weighting is required for \ACP and \aPTodd, since these observables involve only one decay mode.

The asymmetry \Araw is determined from a simultaneous extended maximum likelihood unbinned fit to the \Lb and \Lbbar invariant mass distributions. The \AT and \ATbar asymmetries are determined by means of a simultaneous extended maximum likelihood unbinned fit to the four subsamples defined above.
The signal model for all fits is the sum of two Crystal Ball functions \cite{Skwarnicki:1986xj}, one with a low-mass power-law tail and one with a high-mass tail, and a Gaussian function, all sharing the same peak position. Only the peak position, the total width of the composite function and the overall normalization are free to vary, with all other shape parameters fixed from a fit to simulated decays. The background is modelled by an exponential function. The raw asymmetry \Araw is incorporated in the fit model as
\begin{equation}
N_{\Lbbar} = N_{\Lb} \frac{1-\Araw}{1+\Araw},
\label{eq::Lbbaryield}
\end{equation}
and \DACP is derived from the raw asymmetries measured in the signal and control modes according to Eq.~\ref{eq::DACP}. The asymmetries \AT and \ATbar are included in the fit as
\def\arraystretch{1.8}
\begin{equation}
\begin{array}{cc}
N_{\Lb,\CT>0} = \frac{1}{2} N_{\Lb} (1+\AT), & N_{\Lb,\CT<0} = \frac{1}{2} N_{\Lb} (1-\AT),\\
N_{\Lbbar,-\CTbar>0} = \frac{1}{2} N_{\Lbbar} (1+\ATbar), & N_{\Lbbar,-\CTbar<0} = \frac{1}{2} N_{\Lbbar} (1-\ATbar),\\
\end{array}
\end{equation}
and the observables \ACP and \aPTodd are computed from \AT and \ATbar, which are found to be uncorrelated. Background yields are fitted independently for each subsample, while all the signal shape parameters are shared among the subsamples.

\begin{figure}
\centering
\includegraphics[scale=0.6]{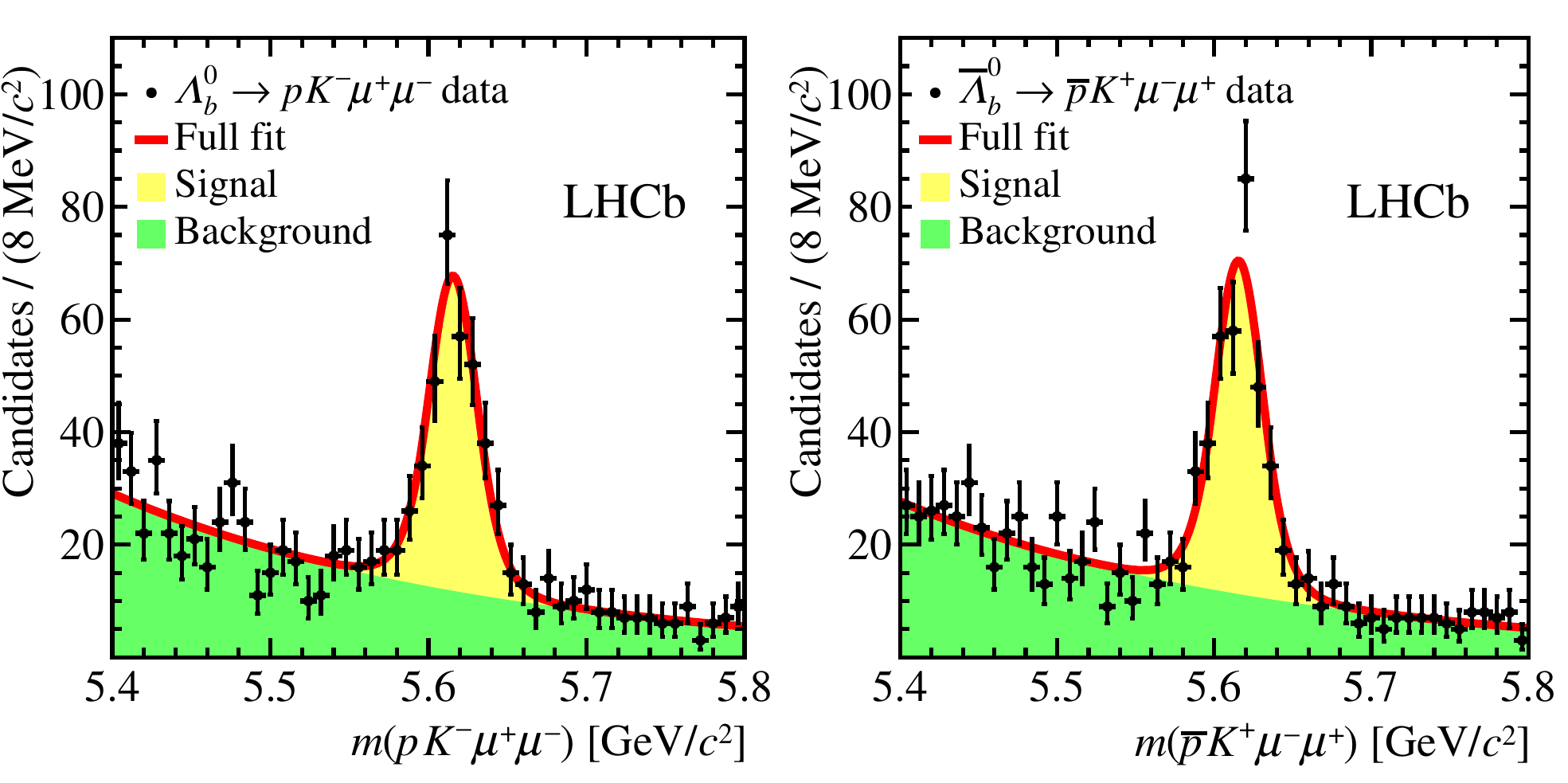}
\includegraphics[scale=0.6]{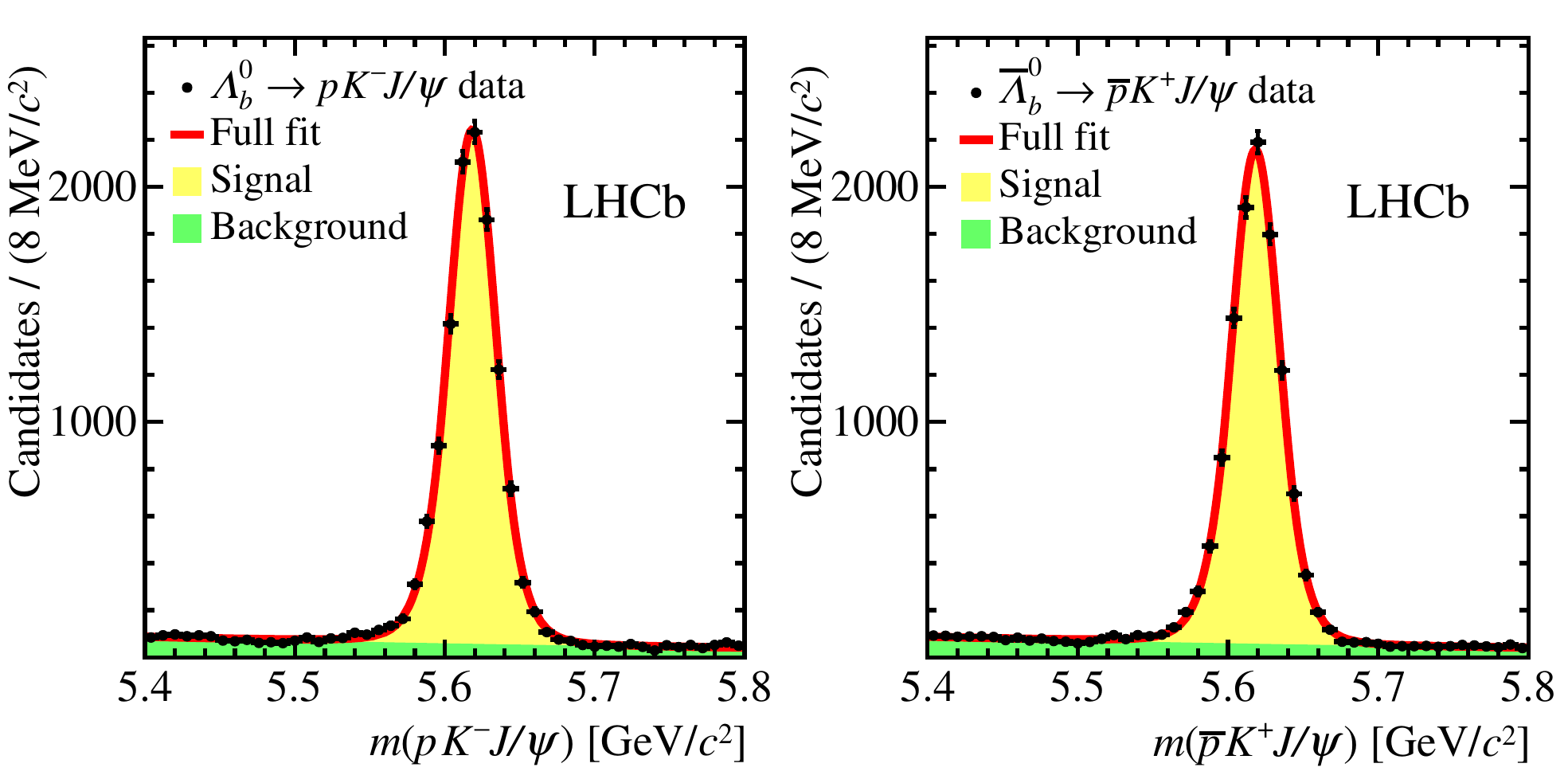}
\caption{Invariant mass distributions of (top) \LK and (bottom) \LKJ candidates, with fit results superimposed. Plots refer to the (left) \Lb and (right) \Lbbar subsamples.\label{fig:ACPfit}}
\end{figure}

The invariant mass distributions of \LK and \LKJ candidates, with fit results superimposed, are shown in Fig.~\ref{fig:ACPfit}.
The \Araw asymmetries are found to be $(-2.8\pm 5.0)\times 10^{-2}$ for signal decays and $(2.0\pm 0.7)\times 10^{-2}$ for the control mode, which yields efficiency-uncorrected $\DACP = (-4.8 \pm 5.0)\times 10^{-2}$. The total signal yields from the fits to the data are $600 \pm 44$ candidates for \LK, and $22\,911\pm 230$ for \LKJ decays. The uncertainties are statistical only. This represents the first observation of the \LK decay mode.

\begin{figure}
\centering
\includegraphics[scale=0.6]{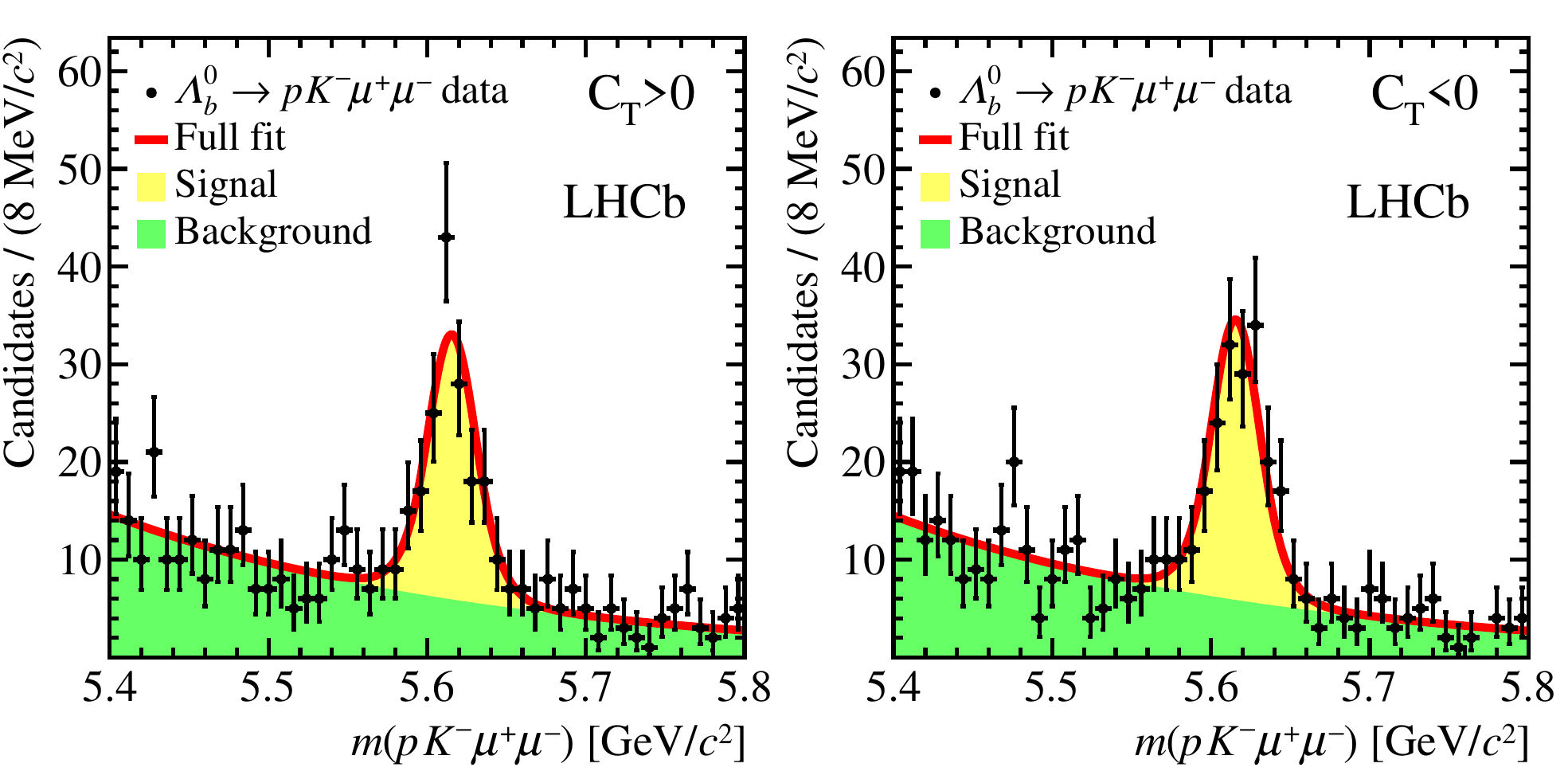}
\includegraphics[scale=0.6]{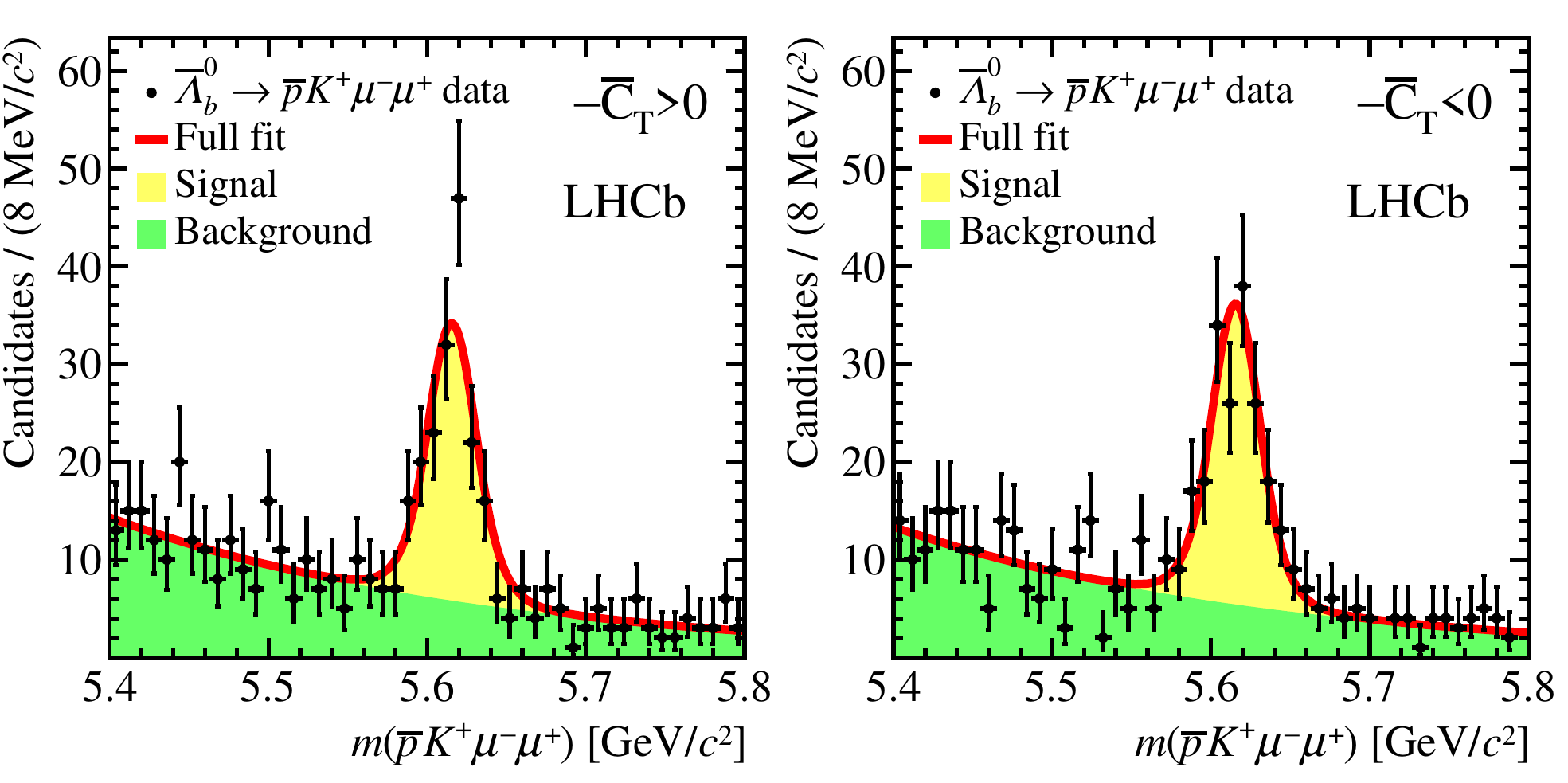}
\caption{Invariant mass distributions of the \LK subsamples used for the \AT and \ATbar measurements. Plots refer to (top) \Lb and (bottom) \Lbbar decays divided into the subsamples (left) $\CT>0,-\CTbar>0$ and (right) $\CT<0,-\CTbar<0$.\label{fig::Toddfit}}
\end{figure}

The invariant mass distributions of the \LK subsamples used for the \AT and \ATbar measurements, with fit results superimposed, are shown in Fig.~\ref{fig::Toddfit}. From the signal yields, the triple-product asymmetries are found to be $\AT = (-2.8\pm 7.2)\times 10^{-2}$ and $\ATbar = (4.0\pm 6.9)\times 10^{-2}$, and the resulting efficiency-uncorrected parity- and \CP-violating observables are
$\aPTodd = (-3.4 \pm 5.0)\times 10^{-2}$ and 
$\ACP = (0.6\pm 5.0)\times 10^{-2}$, where again the uncertainties are statistical only.

\section{Systematic uncertainties}
\label{sec:systematics}
The analysis method depends upon the weighting procedure discussed in Sec.~\ref{sec:asymmetries} to equalise the kinematic distributions of the protons and kaons between the signal and control modes. For \DACP, the associated systematic uncertainty is estimated by varying the weights within their uncertainties and taking the largest deviation, $\pm\; 0.15\times 10^{-2}$, as a systematic uncertainty. No weighting is needed for \ACP and \aPTodd, and therefore no systematic uncertainty is assigned.
Instead, the effects of selection and detector acceptance on the triple-product asymmetries are estimated by measuring $\ACP(p\Km\jpsi)$ on the control mode, \LKJ. A value of $(0.5\pm 0.7)\times 10^{-2}$ is obtained. For this mode negligible \CPV is expected, and the statistical uncertainty of the measured asymmetry is assigned as the corresponding systematic uncertainty on the observables \ACP and \aPTodd.
The effects of the reconstruction efficiency on the measured observables are considered by weighting each event by the inverse of the efficiency extracted from simulated events. This leads to a change in the central values of $+1.3\times 10^{-2}$ on \DACP, of $+0.6\times 10^{-2}$ on \ACP and of $-1.4\times 10^{-2}$ on \aPTodd.
A systematic uncertainty is assigned by varying the efficiencies within their uncertainties. This amounts to $\pm\; 0.10\times 10^{-2}$ for the \DACP observable and to $\pm\; 0.02\times 10^{-2}$ for \ACP and \aPTodd.

The above effects are the dominant sources of systematic uncertainties. Other possible sources of systematic uncertainties are considered. The experimental resolution on \CT is studied with simulated signal events. The effect of the fit model choice is studied by fitting simulated pseudoexperiments with an alternative fit model, in which the Crystal Ball functions are replaced with bifurcated Gaussian functions and the exponential background shape is replaced with a polynomial.
Systematic effects from \Lb polarisation~\cite{LHCb-PAPER-2012-057}, multiple candidates, and residual physical backgrounds are also studied. These contributions have negligible impact on the measured asymmetries.


\section{Conclusions}
The first search for \CP violation in the process \LK is performed with a data sample containing $600\pm 44$ signal decays, this representing the first observation of this \Lb decay mode.
Two different \CP-violating observables that are sensitive to different manifestations of \CP violation, \DACP and \ACP, are measured.
The parity-violating observable \aPTodd is also measured. The values obtained are
\begin{align}
\DACP &=  (-3.5 \pm 5.0\;(\mathrm{stat}) \pm 0.2\;(\mathrm{syst}))\times 10^{-2}, \nonumber\\[1.5ex]
\ACP &= (\phantom{-}1.2 \pm 5.0\;(\mathrm{stat}) \pm 0.7\;(\mathrm{syst}))\times 10^{-2}, \nonumber\\[1.5ex]
\aPTodd &= (-4.8 \pm 5.0\;(\mathrm{stat}) \pm 0.7\;(\mathrm{syst}))\times 10^{-2}.\nonumber
\end{align}
The results are compatible with \CP and parity conservation and agree with SM predictions for \CPV~\cite{Alok:2011gv,Paracha:2014hca}, and with experimental results~\cite{LHCB-PAPER-2014-032,Lees:2012tva} for decays mediated by $b\to s\mu^+\mu^-$ transitions in \Bz and \Bp meson decays.








\section*{Acknowledgements} 
\noindent We express our gratitude to our colleagues in the CERN
accelerator departments for the excellent performance of the LHC. We
thank the technical and administrative staff at the LHCb
institutes. We acknowledge support from CERN and from the national
agencies: CAPES, CNPq, FAPERJ and FINEP (Brazil); MOST and NSFC (China);
CNRS/IN2P3 (France); BMBF, DFG and MPG (Germany); INFN (Italy); 
FOM and NWO (The Netherlands); MNiSW and NCN (Poland); MEN/IFA (Romania); 
MinES and FASO (Russia); MinECo (Spain); SNSF and SER (Switzerland); 
NASU (Ukraine); STFC (United Kingdom); NSF (USA).
We acknowledge the computing resources that are provided by CERN, IN2P3 (France), KIT and DESY (Germany), INFN (Italy), SURF (The Netherlands), PIC (Spain), GridPP (United Kingdom), RRCKI and Yandex LLC (Russia), CSCS (Switzerland), IFIN-HH (Romania), CBPF (Brazil), PL-GRID (Poland) and OSC (USA). We are indebted to the communities behind the multiple open 
source software packages on which we depend.
Individual groups or members have received support from AvH Foundation (Germany),
EPLANET, Marie Sk\l{}odowska-Curie Actions and ERC (European Union), 
Conseil G\'{e}n\'{e}ral de Haute-Savoie, Labex ENIGMASS and OCEVU, 
R\'{e}gion Auvergne (France), RFBR and Yandex LLC (Russia), GVA, XuntaGal and GENCAT (Spain), Herchel Smith Fund, The Royal Society, Royal Commission for the Exhibition of 1851 and the Leverhulme Trust (United Kingdom).

\newpage


\addcontentsline{toc}{section}{References}
\setboolean{inbibliography}{true}
\bibliographystyle{LHCb}
\bibliography{main,LHCb-PAPER,LHCb-CONF,LHCb-DP,LHCb-TDR}
%
%
%

 
\newpage
\centerline{\large\bf LHCb collaboration}
\begin{flushleft}
\small
R.~Aaij$^{40}$,
B.~Adeva$^{39}$,
M.~Adinolfi$^{48}$,
Z.~Ajaltouni$^{5}$,
S.~Akar$^{59}$,
J.~Albrecht$^{10}$,
F.~Alessio$^{40}$,
M.~Alexander$^{53}$,
S.~Ali$^{43}$,
G.~Alkhazov$^{31}$
% ...

{\footnotesize \it
% ...
$ ^{5}$Clermont Universit{\'e}, Universit{\'e} Blaise Pascal, CNRS/IN2P3, LPC, Clermont-Ferrand, France\\
% ...
$ ^{10}$Fakult{\"a}t Physik, Technische Universit{\"a}t Dortmund, Dortmund, Germany\\
% ...
$ ^{31}$Petersburg Nuclear Physics Institute (PNPI), Gatchina, Russia\\
% ...
$ ^{39}$Universidad de Santiago de Compostela, Santiago de Compostela, Spain\\
$ ^{40}$European Organization for Nuclear Research (CERN), Geneva, Switzerland\\
% ...
$ ^{43}$Nikhef National Institute for Subatomic Physics, Amsterdam, The Netherlands\\
% ...
$ ^{48}$H.H. Wills Physics Laboratory, University of Bristol, Bristol, United Kingdom\\
% ...
$ ^{53}$School of Physics and Astronomy, University of Glasgow, Glasgow, United Kingdom\\
% ...
$ ^{59}$University of Cincinnati, Cincinnati, OH, United States\\
% ...
}
\end{flushleft}

%
%
%
%
\end{document}